\documentclass{JAC2000}
\usepackage{graphicx}
\newcommand{\ud}{\mathrm{d}}
\begin{document}
\title{QUASICLASSICAL CALCULATIONS IN BEAM DYNAMICS}
\author{A. Fedorova,  M. Zeitlin, 
IPME, RAS, V.O. Bolshoj pr., 61, 199178, St.~Petersburg, Russia 
\thanks{e-mail: zeitlin@math.ipme.ru}\thanks{ http://www.ipme.ru/zeitlin.html;
http://www.ipme.nw.ru/zeitlin.html}
}

\maketitle

\begin{abstract}
We present some applications of general harmonic/wavelet analysis
approach (generalized coherent states, wavelet pa\-ck\-ets) to 
numerical/analytical calculations in (nonlinear) quasiclassical/quantum
beam dynamics problems.
(Naive) deformation quantization, multiresolution representations
and Wigner transform are the key points.

\end{abstract}

\section{INTRODUCTION}

In this paper we consider some starting points in the applications 
of a new numerical-analytical 
technique which is based on the methods of local nonlinear harmonic 
analysis (wavelet analysis, generalized coherent states analysis) 
to the quantum/quasiclassical
(nonlinear) beam/accelerator physics calculations.
The reason for this treatment is that recently  a number
of problems appeared in which one needs take into account quantum properties of
particles/beams.
We mention only two: diffractive quantum limits of accelerators
(achievable transverse beam spot size) and the description of
dynamical evolution of high density beams by using collective models [1].
Our starting point is the general point of view of deformation
quantization approach at least on naive Moyal/Weyl/Wigner
level (from observables to symbols) (part 2). Then
we present some useful numerical wavelet analysis technique, which gives
the most sparse representation for two main operators
(multiplication and differentiating) in any Hilbert space of states.
Wavelet analysis is a some set of mathematical
methods, which gives us the possibility to work with well-localized bases
(Fig.1) in
functional spaces and gives for the general type of operators (differential,
integral, pseudodifferential) in such bases the maximum sparse forms. 
The approach from this paper is related to our 
investigation of classical nonlinear dynamics 
of accelerator/beam problems [2]-[10].   
The common point is that
any solution
which comes from full multiresolution expansion in all time 
scales gives us expansion into a slow part
and fast oscillating parts. So, we may move
from coarse scales of resolution to the 
finest one for obtaining more detailed information about our dynamical process.
In this way we give contribution to our full solution
from each scale of resolution or each time scale.
The same is correct for the contribution to power spectral density
(energy spectrum): we can take into account contributions from each
level/scale of resolution.
Because affine
group of translations and dilations (or more general group, which acts on the space of solutions) 
is inside the approach
(in wavelet case), this
method resembles the action of a microscope. We have contribution to
final result from each scale of resolution from the whole
infinite scale of spaces. 
Besides affine group symmetry, in part 3 we consider modelling, based on very 
useful and quantum oriented
Wigner transform/function approach (corresponding to Weyl-Heisenberg group), 
which explicitly demonstrates quantum interference
of (coherent) states.
\begin{figure}[ht]
\centering
\includegraphics*[width=60mm]{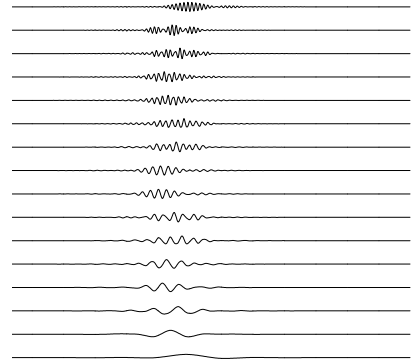}
\caption{Localized contributions to beam motion.}
\end{figure}

\section{Quasiclassical Evolution}                                            
                                                                              
Let us consider classical and quantum dynamics in phase space               
$\Omega=R^{2m}$ with coordinates $(x,\xi)$ and generated by                   
Hamiltonian ${\cal H}(x,\xi)\in C^\infty(\Omega;R)$.                     
If $\Phi^{\cal H}_t:\Omega\longrightarrow\Omega$ is (classical) flow then
time evolution of any bounded classical observable or                      
symbol $b(x,\xi)\in C^\infty(\Omega,R)$ is given by $b_t(x,\xi)=                
b(\Phi^{\cal H}_t(x,\xi))$. Let $H=Op^W({\cal H})$ and $B=Op^W(b)$ are
the self-adjoint operators or quantum observables in $L^2(R^n)$,                
representing the Weyl quantization of the symbols ${\cal H}, b$ [12]   
\begin{eqnarray*}                                                       
&&(Bu)(x)=\frac{1}{(2\pi\hbar)^n}\int_{R^{2n}}b\left(\frac{x+y}{2},\xi\right)
\cdot\\                                                                       
&&e^{i<(x-y),\xi>/\hbar}u(y)\ud y\ud\xi,                             
\end{eqnarray*}                                           
where $u\in S(R^n)$ and $B_t=e^{iHt/\hbar}Be^{-iHt/\hbar}$ be the    
Heisenberg observable or quantum evolution of the observable $B$       
under unitary group generated by $H$. $B_t$ solves the Heisenberg equation of
motion                                                             
$\dot{B}_t=({i}/{\hbar})[H,B_t].$                                        
Let $b_t(x,\xi;\hbar)$ is a symbol of $B_t$ then we have      
 the following equation for it                                        
\begin{equation}                                                             
\dot{b}_t=\{ {\cal H}, b_t\}_M,                                    
\end{equation}                                    
with the initial condition $b_0(x,\xi,\hbar)=b(x,\xi)$.             
Here $\{f,g\}_M(x,\xi)$ is the Moyal brackets of the observables
$f,g\in C^\infty(R^{2n})$, $\{f,g\}_M(x,\xi)=f\sharp g-g\sharp f$,
where $f\sharp g$ is the symbol of the operator product and is presented
by the composition of the symbols $f,g$                           
\begin{eqnarray*}
&&(f\sharp g)(x,\xi)=\frac{1}{(2\pi\hbar)^{n/2}}\int_{R^{4n}}
e^{-i<r,\rho>/\hbar+i<\omega,\tau>/\hbar}\\
&& \cdot f(x+\omega,\rho+\xi)\cdot
g(x+r,\tau+\xi)\ud\rho \ud\tau \ud r\ud\omega.
\end{eqnarray*}
For our problems it is useful that $\{f,g\}_M$ admits the formal
expansion in powers of $\hbar$:
\begin{eqnarray*}
&&\{f,g\}_M(x,\xi)\sim \{f,g\}+2^{-j}\cdot\\
&&\sum_{|\alpha+\beta|=j\geq 1}(-1)^{|\beta|}\cdot
(\partial^\alpha_\xi fD^\beta_x g)\cdot(\partial^\beta_\xi
 gD^\alpha_x f),
\end{eqnarray*}
 where $\alpha=(\alpha_1,\dots,\alpha_n)$ is
a multi-index, $|\alpha|=\alpha_1+\dots+\alpha_n$,
$D_x=-i\hbar\partial_x$.
So, evolution (1) for symbol $b_t(x,\xi;\hbar)$ is
\begin{eqnarray}
&&\dot{b}_t=\{{\cal H},b_t\}+\frac{1}{2^j}
\sum_{|\alpha|+\beta|=j\geq 1}(-1)^{|\beta|}
\cdot\\
&&\hbar^j
(\partial^\alpha_\xi{\cal H}D_x^\beta b_t)\cdot
(\partial^\beta_\xi b_t D_x^\alpha{\cal H}).\nonumber
\end{eqnarray}

At $\hbar=0$ this equation transforms to classical Liouville equation
\begin{equation}
\dot{b}_t=\{{\cal H}, b_t\}.
\end{equation}
Equation (2) plays a key role in many quantum (semiclassical) problem.
Our approach to solution of systems (2), (3) is based on our technique
from [11] and very useful linear parametrization for differential operators
which we present now.
Let us consider multiresolution representation
$\dots\subset V_2\subset V_1\subset V_0\subset V_{-1}
\subset V_{-2}\dots$. Let T be an operator $T:L^2(R)
\rightarrow L^2(R)$, with the kernel $K(x,y)$ and
$P_j: L^2(R)\rightarrow V_j$ $(j\in Z)$ is projection
operators on the subspace $V_j$ corresponding to j level of resolution:
$(P_jf)(x)=\sum_k<f,\varphi_{j,k}>\varphi_{j,k}(x).$ Let
$Q_j=P_{j-1}-P_j$ is the projection operator on the subspace $W_j$ then
we have the following "microscopic or telescopic"
representation of operator T which takes into account contributions from
each level of resolution from different scales starting with
coarsest and ending to finest scales [13]:
$
T=\sum_{j\in Z}(Q_jTQ_j+Q_jTP_j+P_jTQ_j).
$
We remember that this is a result of presence of affine group inside this
construction.
The non-standard form of operator representation [13] is a representation of
an operator T as  a chain of triples
$T=\{A_j,B_j,\Gamma_j\}_{j\in Z}$, acting on the subspaces $V_j$ and
$W_j$:
$
 A_j: W_j\rightarrow W_j, B_j: V_j\rightarrow W_j,
\Gamma_j: W_j\rightarrow V_j,
$
where operators $\{A_j,B_j,\Gamma_j\}_{j\in Z}$ are defined
as
$A_j=Q_jTQ_j, \quad B_j=Q_jTP_j, \quad\Gamma_j=P_jTQ_j.$
The operator $T$ admits a recursive definition via
$$T_j=
\left(\begin{array}{cc}
A_{j+1} & B_{j+1}\\
\Gamma_{j+1} & T_{j+1}
\end{array}\right),$$
where $T_j=P_jTP_j$ and $T_j$ works on $ V_j: V_j\rightarrow V_j$.
It should be noted that operator $A_j$ describes interaction on the
scale $j$ independently from other scales, operators $B_j,\Gamma_j$
describe interaction between the scale j and all coarser scales,
the operator $T_j$ is an "averaged" version of $T_{j-1}$.
We may compute such non-standard representations of operator $\ud/\ud x$ in the
wavelet bases by solving only the system of linear algebraical
equations. 
Let $r_\ell=\int\varphi(x-\ell)\frac{\ud}{\ud x}\varphi(x)\ud x, \ell\in Z.$
Then, the representation of $d/dx$ is completely determined by the
coefficients $r_\ell$ or by representation of $d/dx$ only on
the subspace $V_0$. 
The coefficients $r_\ell, \ell\in Z$ satisfy the
usual system of linear algebraical equations.
For the representation of operator $d^n/dx^n$ we have the similar reduced
linear system of equations.
Then finally we have for action of operator $T_j(T_j:V_j\rightarrow V_j)$
on sufficiently smooth function $f$:
$$
(T_j f)(x)=\sum_{k\in Z}\left(2^{-j}\sum_{\ell}r_\ell f_{j,k-\ell}\right)
\varphi_{j,k}(x),
$$
where $\varphi_{j,k}(x)=2^{-j/2}\varphi(2^{-j}x-k)$ is wavelet basis and
$$
f_{j,k-1}=2^{-j/2}\int f(x)\varphi(2^{-j}x-k+\ell)\ud x
$$
are wavelet coefficients. So, we have simple linear para\-met\-rization of
matrix representation of our differential operator in wavelet basis
and of the action of
this operator on arbitrary vector in our functional space.
Then we may use such representation in all quasiclassical calculations.

\section{Wigner Transform}

According to Weyl transform (observable-symbol) state or wave function corresponds to
Wigner function, which is analog of classical phase-space distribution.
If $\psi(x,t), x\in R^n$ satisfies the Schroedinger equation
\begin{equation}
i\hbar \partial_t\psi=-(\hbar^2/2)\triangle\psi+V\psi
\end{equation}
and W is the Wigner transform of $\psi$
\begin{eqnarray} 
W(t,x,\nu)=&&\int e^{-i\nu y}\bar\psi(t,x+(\hbar/2)y)\cdot \nonumber\\
 &&\psi(t,x-(\hbar/2)y)\ud y,
\end{eqnarray}
then W satisfies the pseudo-differential ($\psi$DO)
Wigner equation
\begin{equation}
\partial_t W+\upsilon\partial_xW-(i/\hbar)P(V)W=0,
\end{equation}

where $\psi$DO operator $P(V)$ is
\begin{eqnarray}
 &P(V)f(x,\nu)=\frac{1}{(2\pi)^n}\int e^{-i\nu y}\Big[
V(x+\frac{\hbar}{2}y)- \nonumber\\
 &V(x-\frac{\hbar}{2}y)\Big]\cdot 
\Big(\int e^{iy\xi}f(x,\xi)\ud\xi\Big)\ud y 
\end{eqnarray}
In quasiclassical limit $\hbar\to 0$ the operator $P(V)$ converges to 
$-\partial_x V\cdot\partial_\nu$. We consider it in [11].
On Fig.~2 we present calculations [14] of Wigner transform for beam motion, represented
by four gaussians, which explicitly demonstrates quantum inteference in the phase space.

We give more details in [11].

\begin{figure*} [htb]
\centering
\includegraphics*[width=150mm]{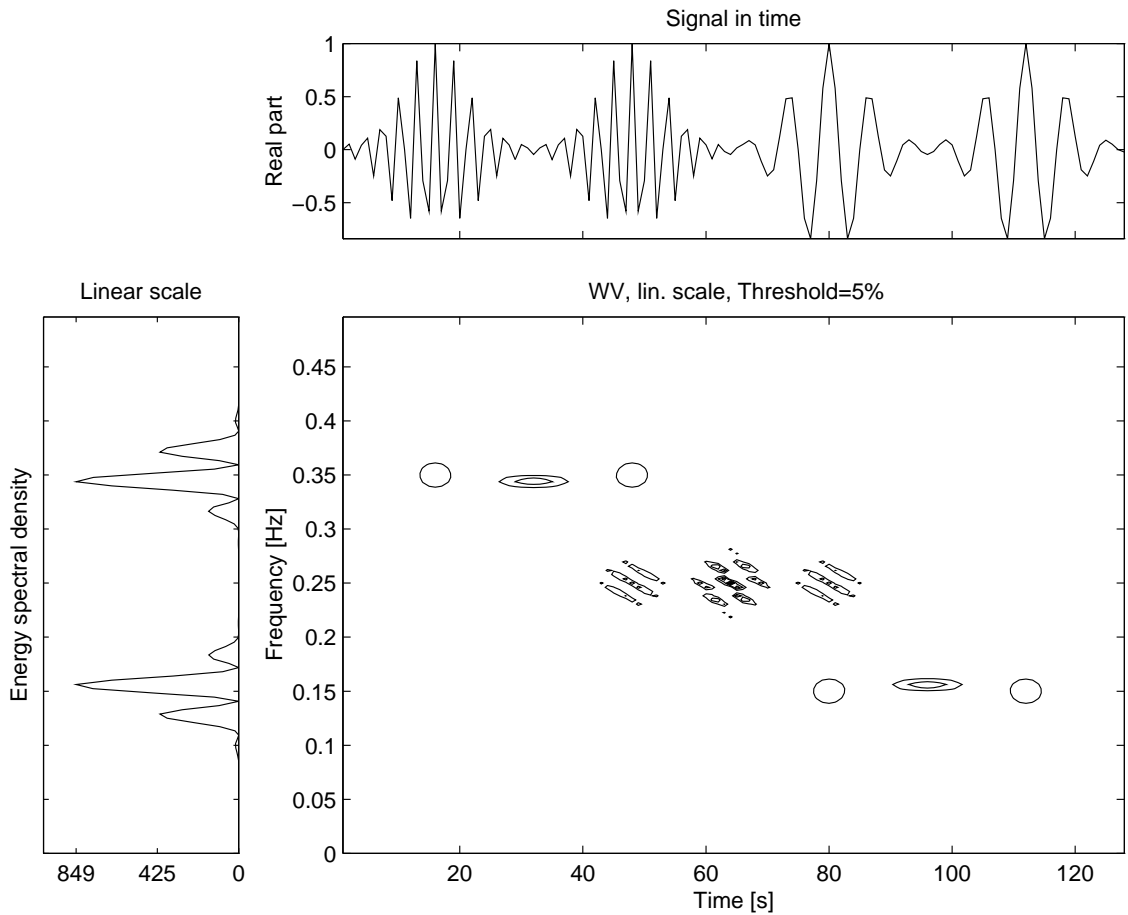}
\caption{Wigner transform }
\end{figure*}

We would like to thank Professor
James B. Rosenzweig and Mrs. Melinda Laraneta for
nice hospitality, help and support during UCLA ICFA Workshop.

 \end{document}